\documentclass[10pt,prd,superscriptaddress,amsfonts,amssymb,amsmath,showpacs,twocolumn]{revtex4-2}
\usepackage{bm}
\usepackage{amsfonts}
\usepackage{latexsym}
\usepackage{graphicx}
\usepackage{amsmath}
\usepackage{palatino}
\usepackage{mathpazo}
\usepackage{textcomp}
\usepackage{float}
\usepackage{booktabs}
\usepackage{dcolumn}
\usepackage{booktabs}
\usepackage{multirow}
\usepackage{hyperref}
\usepackage{orcidlink}
\hypersetup{colorlinks,citecolor=blue}
\usepackage{xcolor}

\begin{document}
	
	\color{black}       

	\title{Cosmological Implications of $f(R,\Sigma, T)$ Gravity: A Unified Approach Using OHD and SN Ia Data}


\author{ N. Myrzakulov\orcidlink{0000-0001-8691-9939}}\email{nmyrzakulov@gmail.com}
\affiliation{L N Gumilyov Eurasian National University, Astana 010008, Kazakhstan}

\author{ S. H. Shekh\orcidlink{0000-0003-4545-1975}}\email{  da\_salim@rediff.com}
\affiliation{L N Gumilyov Eurasian National University, Astana 010008, Kazakhstan}
\affiliation{Department of Mathematics, S.P.M. Science and Gilani Arts, Commerce College, Ghatanji, Yavatmal, \\Maharashtra-445301, India.}

\author{Anirudh Pradhan\orcidlink{0000-0002-1932-8431}}\email{pradhan.anirudh@gmail.com}
\affiliation{Centre for Cosmology, Astrophysics and Space Science (CCASS)
	GLA University, Mathura 281406, Uttar Pradesh, India}

	\begin{abstract}
		\textbf{Abstract:} This paper investigates the cosmological implications of the modified gravity framework known as $f(R,\Sigma, T)$ gravity, focusing on its potential to unify and extend current cosmological models. The theory, introduced by Bakry and Ibraheem in 2023, combines the Ricci scalar, a scalar parameter  representing torsion or other geometric properties, and the trace of the energy-momentum tensor. By analyzing observational Hubble data (OHD) and the Pantheon compilation of Type Ia Supernovae (SN Ia), we explore how this framework provides the accelerated expansion of the universe, the nature of dark energy, and phenomena like the Big Rip singularity. Employing the Friedmann-Robertson-Walker (FRW) metric and solving the modified field equations, we derive key cosmological parameters such as the Hubble constant  and matter energy density parameter. These parameters are constrained through statistical analysis of observational data, yielding  and  from OHD, and  and  from SN Ia. The evolution of the equation of state (EoS) parameter, isotropic pressure, and energy density is also investigated. By considering energy conditions and stability criteria, the study highlights the viability of  gravity as an alternative framework to General Relativity and  models. Our findings affirm the model’s compatibility with current observational evidence and its potential to the universe’s past and future dynamics.
	\end{abstract}

	\maketitle
	
	\section{Introduction}
	As, in 1998, two independent groups of astronomers made groundbreaking discoveries about the universe's expansion. The High-z Supernova Search Team, led by Riess et al. \cite{1}, and the Supernova Cosmology Project, headed by Perlmutter et al. \cite{2}, focused their studies on a particular type of exploding star known as supernovae. Their goal was to determine whether the universe's expansion was slowing down or speeding up. Surprisingly, both teams discovered that the gravitational force of matter is causing the universe's expansion to accelerate rather than decelerate. This unexpected result revolutionized cosmology, leading to the concept of \textit{dark energy}, an unknown force responsible for the accelerated expansion. \textit{dark energy} is characterized by negative pressure and behaves like anti-gravity. Despite extensive research, its exact nature and properties remain unknown. To learn more about this mysterious occurrence, scientists are still investigating different possibilities and carrying out observational research. According to Planck's observations, \textit{dark energy} is thought to make up roughly 68.3\% of the universe \cite{3,4}.\\
	This mysterious \textit{dark energy} and the accelerated expansion of the universe have also spurred interest in exploring alternative theoretical frameworks beyond the general theory of relativity called \textit{Modified theories of gravity} which  provide alternative frameworks that aim to extend or revise the general theory of relativity to address phenomena that remain unexplained by relativity alone. These theories seek to offer a more comprehensive understanding of the universe's nature and dynamics. By introducing innovative concepts and mathematical models, they strive to refine our knowledge of gravity and its interactions with matter and space-time. Examples include $f(R)$ gravity \cite{5,6,6a,6b,6c}, where $R$ represents the Ricci scalar; $f(T)$ gravity \cite{7,7a,7b,7c,7d,7e,7f,7g,7h}, where $T$ denotes the torsion scalar; $f(G)$ gravity \cite{8,8a,8b}, with $G$ being the Gauss-Bonnet term; $f(R, T)$ gravity \cite{9}, involving both the Ricci scalar $R$ and the trace of the energy-momentum tensor $T$; $f(R, G)$ gravity \cite{10,11,11a}, combining the Ricci scalar and the Gauss-Bonnet term; $f(Q)$ gravity \cite{12,12a,12b,12c,12d,12e}, where $Q$ is the non-metricity scalar; $f(Q, T)$ gravity \cite{13,13a,13b,13c}, which includes both the non-metricity scalar $Q$ and the trace of the energy-momentum tensor; and The $f(R,\Sigma, T)$ \cite{14} gravity  model is among the advanced theoretical frameworks that investigate the accelerated expansion of the universe, encompassing both its early and late evolutionary stages. Introduced by Bakry and Ibraheem in 2023 \cite{14}, this theory is rooted in the concept of Absolute Parallelism (AP) geometry. It examines the interplay between normal gravity, strong gravity, and antigravity by adjusting key parameters in their formulation. 
	
	
	Among these, the \( f(R,\Sigma,T) \) gravity theory has emerged as a promising alternative. This framework extends the conventional \( f(R) \) and \( f(R,T) \) models by introducing an additional scalar parameter \( \Sigma \), which may represent torsion or other geometric properties of space-time. By coupling the Ricci scalar \( R \), the scalar \( \Sigma \), and the trace of the energy-momentum tensor \( T \), this theory provides a unified platform for studying the accelerated expansion of the universe and \textit{dark energy}. Moreover, it offers valuable insights into phenomena such as the Big Rip singularity and enables exploration of gravitational interactions beyond the scope of general relativity.

	The action principal to derive a set of field equations is defined as
	\begin{small}
		\begin{equation}\label{e1}
			I = \frac{1}{16\pi} \int_{\Omega} \sqrt{-g} \left( B + \mathcal{L}_m \right) d^4x = \frac{1}{16\pi} \int_{\Omega} \sqrt{-g} \left( g^{ab} B_{ab} + \mathcal{L}_m \right) d^4x.
		\end{equation}
	\end{small}
	where $\Omega$ is a region enclosed within some closed surface, $B_{ab}=R_{xy}\left\{ \right\}+\Sigma (\Sigma_{xy})$ and $\Sigma_{xy} = b \Phi(xy)$, $b$ be the parameter which is the ratio between gravity and antigravity (i.e. attraction and repulsion) within a given system. The action principle can be expressed in the following manner
	
	\begin{equation}\label{e2}
		I = \frac{1}{16\pi} \int_{\Omega} \sqrt{-g} \left( f(R, \Sigma, T) + \mathcal{L}_m \right) d^4x.
	\end{equation}
	
	By varying the gravitational action (\ref{e2}) with respect to the metric tensor, we can arrive at the field equation for $f(R, \Sigma, T)$  Gravity as follows
	\begin{widetext}
		\begin{equation}\label{e3}
			R_{ab} \frac{\partial f}{\partial R} + R_{ab} \frac{\partial f}{\partial R} - \frac{1}{2} g_{ab} f + \left( g_{ab} \nabla^\gamma \nabla_\gamma - \nabla_a \nabla_b \right) \frac{\partial f}{\partial R} - \beta \frac{\partial f}{\partial R} \\
			= 8\pi T_{ab} +\left( T_{ab} + p g_{ab}\right) \frac{\partial f}{\partial T},
		\end{equation}
	\end{widetext}
In this study, we extend the modified gravity framework by considering the \( f(R,\Sigma,T) \) gravity model, where \(\Sigma \) represents an additional geometric parameter that can encapsulate torsional effects. While teleparallel gravity typically formulates gravity using the Weitzenböck connection and considers the vierbein as the fundamental field, our approach integrates both curvature and torsion contributions in a more generalized manner. This formulation allows for a unified treatment of gravitational interactions, accommodating elements from both metric-affine and teleparallel theories.
To clarify the role of torsion in our framework, we explicitly define  \(\Sigma \) as a term that captures deviations from standard curvature-based formulations, incorporating possible torsional effects. Unlike purely teleparallel models, where the Weitzenböck connection ensures zero curvature while torsion is responsible for gravitational interactions, our model retains the flexibility to describe gravitational dynamics through a combination of curvature and torsion-like influences. This makes it a broader generalization that can recover different gravitational paradigms under specific limits.

	The energy–momentum tensor for matter is defined as
	\begin{equation}\label{e4}
		T_{ab} = (\rho + p) u_a u_b - p g_{ab},
	\end{equation}
	where $\rho$ is the energy density of matter, $p$ is
	the pressure of the matter, and $u_a$ is the fluid-four velocity vector, where $u_a u^b =1$. In this article, we assume that the function $f(R, \Sigma, T)$ is
	given by
	\begin{equation}\label{e5}
		f(R, \Sigma, T) = R + \Sigma + 2\pi \eta T.
	\end{equation}
	where $\eta$ is an arbitrary constant parameter. Using the function (\ref{e5}), the final form of $f(R, \Sigma, T)$ equation (\ref{e3}) is given by
	\begin{equation}\label{e6}
		R_{ab} + \Sigma_{ab} - \frac{1}{2} g_{ab} (R + \Sigma) = 2(4\pi + \pi \eta) T_{ab} + \pi \eta g_{ab}(T + 2p)
	\end{equation}
	which can be alternatively expressed as
	\begin{small}
		\begin{equation}\label{e7}
			B_{ab} = R_{ab} + \Sigma_{ab} = 2(4\pi + \pi \eta) \left(T_{ab} - \frac{1}{2} g_{ab} T\right) - \pi \eta g_{ab} (T + 2p).
		\end{equation}
			\end{small}
		Here the arbitrary constant $\eta$ governs the coupling strength between the matter-energy content and the geometric curvature, influencing the dynamics of the universe described by the field equations. The value of $\eta$ affects how the curvature of spacetime responds to matter and \textit{dark energy}, thereby altering the evolution of the universe's expansion. When \(\eta = 0\), the function reduces to \(f(R, \Sigma, T) = R +\Sigma\), decoupling the matter contribution and simplifying the dynamics to a purely geometric modification of gravity. For \(\eta > 0\), the positive coupling enhances the interaction between matter and curvature, leading to accelerated cosmic expansion and stronger \textit{dark energy} effects. In contrast, \(\eta < 0\) introduces a negative coupling, which could decelerate the universe’s expansion or yield alternative behaviors for \textit{dark energy}, potentially affecting the fate of the universe, such as delaying or intensifying the Big Rip scenario. These variations make \(\eta\) a pivotal parameter in understanding different cosmological evolutions. Hence for the further study we choose \(\eta > 0\) because this choice is particularly suited for investigating scenarios like the Big Rip, where the universe undergoes an exponential acceleration. By adopting \(\eta > 0\), the model captures the essential dynamics of accelerated expansion while providing a framework to explore the intricate interplay between matter, curvature, and \textit{dark energy}. 
		Furthermore, recent studies have critically examined the robustness of the $\Lambda$CDM model in light of new observational data such as Odintsov et al. \cite{14a} introduced a generalized exponential $f(R)$ gravity model and compared it to the $\Lambda$CDM framework using the latest datasets, including the Pantheon+ compilation of Type Ia supernovae, Hubble parameter measurements, cosmic microwave background data, and baryon acoustic oscillations from the DESI collaboration. Their findings suggest that the standard exponential $f(R)$ models provide a significantly better fit than the $\Lambda$CDM model, which is excluded at a 4$\sigma$ confidence level. Additionally, their analysis indicates a non-constant equation of state parameter for dark energy, further challenging the $\Lambda$CDM paradigm whereas Colgain et al. \cite{14b} in their study analyze data from the Dark Energy Spectroscopic Instrument (DESI) to assess the consistency of the $\Lambda$CDM model. They identify a 2$\sigma$ discrepancy in the Luminous Red Galaxy sample at an effective redshift of 0.51, leading to an unexpectedly high matter density parameter, $\Omega_m = 0.668$. This anomaly suggests a preference for a dark energy equation of state parameter $\omega_0 >-1$, indicating potential deviations from the standard $\Lambda$CDM model. The authors emphasize the necessity of understanding the underlying causes of this discrepancy, as it may have significant implications for our comprehension of cosmic acceleration and the nature of \textit{dark energy}.
		\section{Metric and the components of equation}
	To facilitate the solution of field equations in $f(R, \Sigma, T)$ extended symmetric teleparallel gravity, simplifying assumptions are often necessary. The Friedmann-Robertson-Walker (FRW) metric, which is homogeneous, isotropic, and spatially flat, is used in this work and is provided by
	
	\begin{equation}\label{e8}
		ds^{2}=-dt^{2}+\delta_{ij} g_{ij} dx^{i} dx^{j},{\;\;\;\;} i,j=1,2,3,.....N .
	\end{equation}
	This choice of metric allows us to explore the cosmological implications of $f(R, \Sigma, T)$ gravity in a straightforward and analytically tractable manner.
	where $g_{ij}$ are the function of $(-t, x^{1}, x^{2}, x^{3})$ and $t$ refers to the cosmological/cosmic time measure in Gyr. In the four-dimensional FRW space-time, the equation above yields the following:
	\begin{equation}\label{e9}
		\delta_{ij} g_{ij}=a^{2}(t),
	\end{equation} 
where $t$ is the cosmic time in Gyr and $a$ is the universe's average scale factor. All three metrics are equal in the FRW universe, as shown by the correlations indicated above ($g_{11} = g_{22} = g_{33} =a^{2}(t)$). \\
	By utilizing Equation (\ref{e8}) for the space–time in a comoving coordinate system, we can obtain the components of $f(R, \Sigma, T)$ gravity field Equations from the equation (\ref{e7}) as follows:
	\begin{equation}\label{e10}
		3(b-1)\left(\dot{H}+H^2\right)=4 \pi \rho+4 \pi \left(3+\eta\right)p
	\end{equation}
	\begin{equation}\label{e11}
		3(1-b)^2 H^2= \pi \left(8+3\eta\right) \rho - \pi \eta p
	\end{equation}
	Hence, the solutions of the differential Eqs. (\ref{e10}) and (\ref{e11}) are given by:
	\begin{equation}\label{e12}
		p = \frac{1-b}{4(8\pi + 6\eta +\pi \eta^2)} \left[ (8+3\eta) \dot{H} + (4(3-b) + 3\eta)H^2 \right]
	\end{equation}
	\begin{equation}\label{e13}
		\rho = \frac{1-b}{4(8\pi^2 + 6\pi \eta + \eta^2)} \left[ -\eta \dot{H} + (12\pi(1-b) + \eta(3 - 4b))H^2 \right]
	\end{equation}
	The system's dynamics are described by two key equations, namely the modified field equations (\ref{e12}) and (\ref{e13}). Since these equations involve three unknowns, an additional equation is needed to fully solve the system. To address this, researchers commonly adopt a model-independent approach, often employing parameterization schemes for cosmological parameters. Various intriguing \textit{dark energy} and modified gravity models emerge depending on how these parameters, such as the equation of state parameter $\omega(z)$ \cite{Eric} or the deceleration parameter $q(z)$ \cite{Cunha}, are parameterized. Notable parameterizations include the Chevalier-Polarski-Linder (CPL) \cite{Chevalier} and Jassal-Bagla-Padmanabhan (JBP) \cite{Jassal} forms for $\omega(z)$. A broader perspective suggests that other geometric or physical parameters can also be parameterized. This approach is significant as it provides solutions to several key challenges in cosmology while reconstructing the Universe's past and future evolution.
	\section{Specific $H(z)$ and Some Cosmographic Parameters} 
	
	To solve the field equations and study the evolution of cosmological parameters over time, we assume that the Universe is filled with a perfect fluid. The expansion rate is characterized using the dimensionless function $E(z)$, defined as:  
	$$E(z) = \frac{H^2(z)}{H_0^2} = \Omega_{0m}(1+z)^3 + \Omega_{0_{Q,T}}$$  
	where $H(z)$ denotes the Hubble parameter at redshift $z$, $H_0$ is the present-day Hubble constant, and $\Omega_{0m}$ is the present-day matter energy density parameter. The term $\Omega_{0_{Q,T}}$ accounts for the energy density parameter derived from the geometry of $f(Q,T)$ gravity. Another form of $E(z)$ can be written as \cite{Lohakare}:  
	$$E(z) = \Omega_{0m}(1+z)^3 + \alpha(1+z)^2 + \beta(1+z) + \mu$$  
	At $z = 0$, the Hubble parameter equals its current value, $H(z) = H_0$, implying $E(z) = 1$. This leads to the constraint $\alpha + \beta + \mu = 1 - \Omega_{0m}$. To satisfy this, we adopt a simple assumption: $\Omega_{0_{Q,T}} = 1 - \Omega_{0m}$. Substituting this, the equation simplifies to:  
	\begin{equation}
		H(z) = H_0 \left[\Omega_{0m}(1+z)^3 + (1 - \Omega_{0m})\right]^{\frac{1}{2}}
	\end{equation}  
	This equation expresses the Hubble parameter $H(z)$ in terms of redshift $z$, the present-day matter energy density parameter $\Omega_{0m}$, and the Hubble constant $H_0$. This approach is designed to effectively capture the Universe's expansion dynamics, particularly the transition from a decelerating to an accelerating phase. This form is chosen to offer a straightforward yet accurate representation of the Hubble parameter, ensuring consistency with observational evidence. In this study, we constrained both  $H_0$ and $\Omega_{0m}$ using the same datasets previously utilized by Irfan et al., which include a combination of the Supernova Pantheon sample, cosmic chronometer data, and other observational data as detailed below.
	
	\subsection{Observational Constraints} In this section, we outline the observational $H(z)$ data (OHD) and the Pantheon compilation of Supernova Type Ia (SN Ia) data, along with the statistical methodology used to constrain the model parameters $H_0$ and $\Omega_{m0}$.
	The $\chi^{2}$ for statistical methodologyis read as 
	\begin{equation}
		\label{chi1}
		\chi^{2} = \sum_{i=1}^{N}\left[\frac{E_{th}(z_{i})- E_{obs}(z_{i})}{\sigma_{i}}\right]^{2}
	\end{equation}
	where $E_{th}(z_{i})$ and $E_{obs}(z_{i})$ denote the theoretical values and observed values of corresponding parameters respectively. $\sigma_{i}$ and N are standard errors in $E_{obs}(z_{i})$ and number of data points.\\
	\subsubsection{Date sets and results}
\begin{figure}[ht]
		\centering
		\includegraphics[width=6.5cm,height=6.5cm,angle=0]{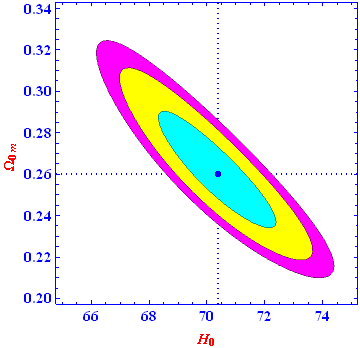}
		\caption{Two-dimensional contour plots at $1\sigma$, $2\sigma$, and $3\sigma$ confidence levels are obtained by constraining our model using the $H(z)$ data set. The unit of $H_{0}$ is $\;km\;s^{-1}\;Mpc^{-1}$}.\label{F1}
	\end{figure}

	\begin{itemize}
		\item {\bf OHD}: We utilized 55 $H(z)$ observational data points covering the redshift range $0 \leq z \leq 2.36$, obtained using the cosmic chronometric technique. These 55 data points are compiled in Table II of Ref. \cite{Lohakare/2022}.
	\end{itemize}
	Fig. \ref{F1} shows the two-dimensional contour plots for this model at $1\sigma$, $2\sigma$, and $3\sigma$ confidence levels based on the observational Hubble data. From this analysis, we estimate the Hubble constant as $H_{0} = 70.37^{+0.84}_{-0.92}$ km/sec/Mpc and the matter energy density parameter as $\Omega_{0m} = 0.26^{+0.015}_{-0.010}$ using the observational $H(z)$ data set.

\begin{figure}[ht]
		\centering
		\includegraphics[width=6.5cm,height=6.5cm,angle=0]{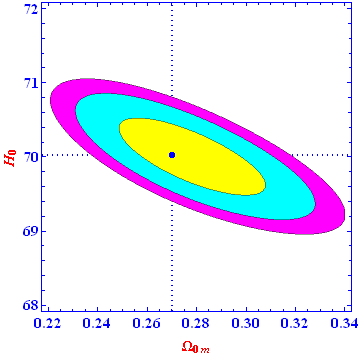}
		\caption{Two-dimensional contour plots at $1\sigma$, $2\sigma$, and $3\sigma$ confidence levels are obtained by constraining our model using the Pantheon compilation of SN Ia data. The unit of $H_{0}$ is $\;km\;s^{-1}\;Mpc^{-1}$}.\label{F2}
	\end{figure}

	\begin{itemize}
		\item {\bf SN Ia}: The Pantheon compilation of SN Ia data \cite{Scolnic/2018} was used, which includes 1048 apparent magnitude measurements in the redshift range $0.01 < z < 2.3$. Additionally, this dataset features 40 binned data points spanning the redshift range $0.0014 \leq z \leq 1.62$ \cite{Scolnic/2018}. The 1048 measurements of apparent magnitude $m_B$ support the redshift range $0.01 \leq z \leq 2.3$.
	\end{itemize}
	Fig. \ref{F2} shows the two-dimensional contour plots for this model at $1\sigma$, $2\sigma$, and $3\sigma$ confidence levels based on the observational SN Ia data set. From this analysis, we estimate the Hubble constant as $H_{0} = 70.02^{+0.44}_{-0.25}$ km/sec/Mpc and the matter energy density parameter as $\Omega_{0m} = 0.27^{+0.025}_{-0.014}$ using the observational SN Ia data set.

	To ensure consistency and simplify further analysis, we adopted the mean value of the two results. This averaged value provides a balanced estimate that incorporates constraints from both datasets and serves as a reliable input for subsequent cosmological analyses, such as modeling the universe's expansion dynamics and evaluating the properties of the Universe.

	\section{Physical behavior}
	We now turn our attention to the physical behavior of the parameters described in equations (\ref{e12}) and (\ref{e13}). From equation (\ref{e13}), the energy density for the model is derived as:  
	\begin{equation}\label{16}
		\begin{aligned}
			\rho& =	\frac{(b+1)}{4 (\eta +2 \pi ) (\eta +4 \pi )} \Bigg\{ H_0^2 ((3-4 b) \eta +12 \pi  (b+1)) \times \\
			& (\Omega_{0m} z (z (z+3)+3)+1)+\frac{3}{2} H_0^2 \Omega_{0m} \eta  (z+1)^3\Bigg\}
		\end{aligned}
	\end{equation}
	
\begin{figure}[ht]
		\centering
		\includegraphics[scale=0.7]{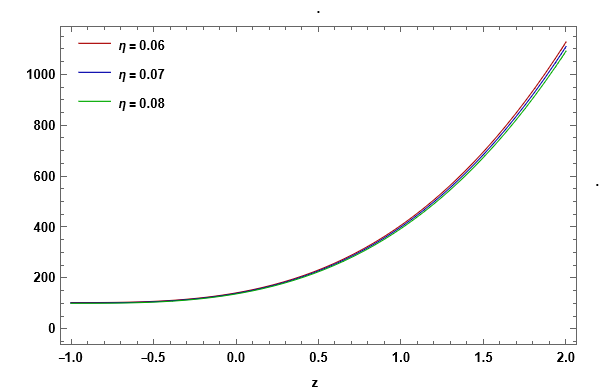}
		\caption{$\rho$ versus $z$}.\label{den}
	\end{figure}

	The plot in Fig. \ref{den} shows the evolution of the energy density \(\rho\) as a function of redshift \(z\) within the framework of \(f(R, \Sigma, T)\) gravity theory. The mathematical formulation for \(\rho\) is provided in equation (38). The figure features three curves—red, blue, and green—corresponding to different values of \(\eta\): \(0.06\), \(0.07\), and \(0.08\), respectively.
	
\begin{figure}[ht]
		\centering
		\includegraphics[scale=0.7]{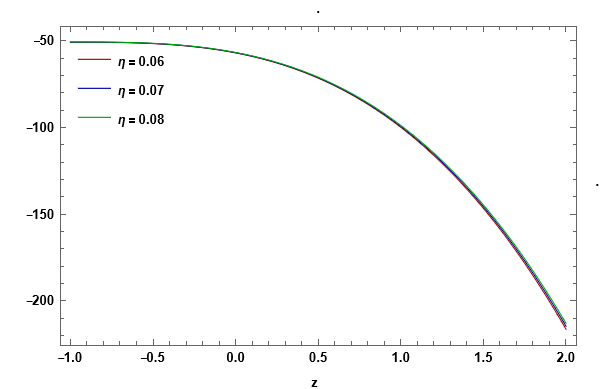}
		\caption{$p$ versus $z$}.\label{p}
	\end{figure}

	Additionally, Fig. \ref{den} illustrates the behavior of the energy density \(\rho\), which shows a steady increase with cosmic redshift \(z\). This trend aligns with the expected behavior in an expanding universe, where energy density increases as the universe contracts and decreases during its expansion. The higher energy density at larger redshifts reflects the conditions of the early universe, while the gradual decrease in \(\rho\) at lower redshifts represents the ongoing dilution of energy density due to cosmic expansion. Such results are consistent with theoretical predictions, reinforcing the reliability of our model in capturing the dynamics of energy density evolution in an expanding universe. These findings further validate the applicability of \(f(R, \Sigma, T)\) gravity in explaining the observed behavior of energy density over cosmic time.\\
	
	and from equation (\ref{e12}), the isotropic pressure for the model is derived as: 
	\begin{equation}\label{17}
		\begin{aligned}
			p &= \frac{(b+1) H_0^2 }{8 (\eta +2 \pi ) (\eta +4 \pi )} \Bigg\{-8 b (\Omega_{0m} z (z (z+3)+3)+1)\\
			&-3 \Omega_{0m} (\eta  (z (z (z+3)+3)+3)+8)+6 (\eta +4)\Bigg\}
		\end{aligned}
	\end{equation}
	The plot in Fig. \ref{p} shows the evolution of the isotropic pressure \(p\) as a function of redshift \(z\) within the framework of \(f(R, \Sigma, T)\) gravity theory. The mathematical formulation for \(\rho\) is provided in equation (38). The figure features three curves—red, blue, and green—corresponding to different values of \(\eta\): \(0.06\), \(0.07\), and \(0.08\), respectively.
	The graphical depiction of \(p\) in Fig. \ref{p} exhibits a distinct negative trend, indicating a monotonic decrease over cosmological time. As shown in Fig. \ref{p}, this negative pressure evolution is consistent with its role in driving the accelerating expansion of the universe, a behavior predicted by various cosmological models. The observed negative isotropic pressure points to a repulsive gravitational effect, which facilitates the acceleration of cosmic expansion. This behavior aligns well with standard cosmological models, such as the \(\Lambda\)CDM model, where dark energy is responsible for the observed accelerated expansion. The agreement between our results and theoretical expectations validates the reliability of our findings and offers valuable properties of the  universe's dynamic.

	We now looked more closely at other physical parameters associated with the Universe's energy density and isotropic pressure. These include the EoS parameter, stability parameter, the $\left(\omega-\omega^{\prime}\right)$ plane, and various energy conditions. A thorough analysis of these parameters is essential for gaining insight into the physical characteristics of the Universe.
	
	\subsubsection{The equation of state parameter}
	The equation of state (EoS) parameter, denoted as \( \omega \), is a key quantity that describes the relationship between the pressure \( p \) and the energy density \( \rho \) of a cosmic fluid and it plays a crucial role in understanding the behavior of different components of the Universe. The EoS parameter can take various forms depending on the type of cosmic fluid. For matter, \( \omega = 0 \), representing non-relativistic matter (dust), while for radiation, \( \omega = \frac{1}{3} \), corresponding to relativistic particles. A cosmological constant or \textit{dark energy} component is characterized by \( \omega = -1 \), which causes the accelerated expansion of the Universe. In models involving \textit{dark energy} or modified gravity, the EoS parameter may vary with redshift, \( \omega(z) \), reflecting the evolution of the Universe’s expansion. Recent observational data, including the Pantheon compilation of Type Ia Supernovae (SN Ia) and cosmic chronometer measurements, suggest that the EoS parameter for the Universe is close to \( \omega \approx -1 \), with slight deviations depending on the specific cosmological model and the redshift range considered. For instance, current estimates suggest that \(\omega \) may vary slightly between \( -1 \) and \( -0.9 \) at different epochs, indicating a possible deviation from a pure cosmological constant. These values are obtained from a combination of observational datasets, including SN Ia, cosmic microwave background (CMB) measurements, and large-scale structure surveys, providing valuable constraints on the nature of \textit{dark energy} and the future evolution of the Universe. Mathematically the equation of state parameter is defined as 
	\begin{equation}
	\omega = \frac{p}{\rho}
	\end{equation}
	For values of the equation of state parameter \( \omega \) differing from \( \omega = -1 \), \textit{dark energy} models can be classified into several categories: Quintessence models, where \( \omega > -1 \); Phantom models, where \( \omega < -1 \); $\kappa$-essence models, in which \( \omega \) is dynamic; and Brane cosmology models, where \( \omega \) varies. These classifications reflect distinct evolutionary trajectories for the Universe, offering alternatives to the static \( \Lambda \)CDM model (here we took the values of \(p\) and \(\rho\) from equations (\ref{16}) and (\ref{17})).
	
	\begin{figure}[ht]
		\centering
		\includegraphics[scale=0.7]{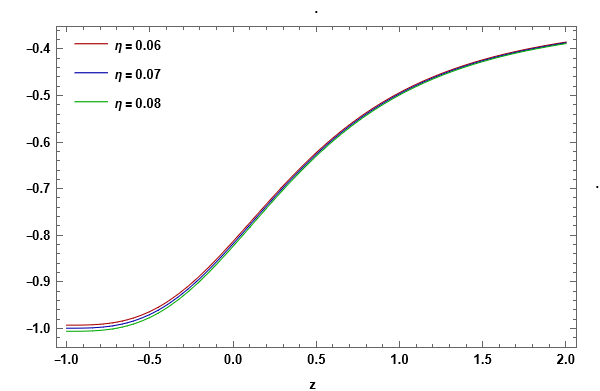}
		\caption{$\omega$ versus $z$}.\label{w}
	\end{figure}  
	The plot in Fig. \ref{w} illustrates the evolution of the equation of state (EoS) parameter \(\omega\) as a function of redshift \(z\) within our model. 
	The figure presents three distinct curves—red, blue, and green—representing different values of \(\eta\): \(0.06\), \(0.07\), and \(0.08\), respectively. 
	
	For \(\eta = 0.06\) (red curve), the universe begins in a matter-dominated phase where \(\omega > 0\). As the universe evolves, \(\omega\) decreases, and the model transitions into a quintessence phase (\(-1 < \omega < 0\)). Notably, this phase persists both in the present and far-future epochs, with \(\omega\) remaining above \(-1\). 
	
	For \(\eta = 0.07\) (blue curve), the universe similarly starts in a matter-dominated phase with \(\omega > 0\). It transitions to a quintessence phase in the present epoch and ultimately approaches the \(\Lambda\)CDM model, converging to \(\omega = -1\) in the far future. This behavior aligns with the cosmological constant scenario.
	
	In contrast, for \(\eta = 0.08\) (green curve), the universe also begins in a matter-dominated phase (\(\omega > 0\)) and transitions to a quintessence phase at present. However, as time progresses, \(\omega\) crosses the phantom divide line (\(\omega = -1\)) and enters a phantom phase where \(\omega < -1\) in the future.
	
	It is noteworthy that for all three values of \(\eta\), the present value of \(\omega\) is approximately \(-0.8\), indicating that the universe is currently in a quintessence phase. The variations in \(\eta\) primarily influence the long-term evolution of \(\omega\), determining whether the universe approaches \(\Lambda\)CDM, remains in the quintessence phase, or transitions into a phantom phase.

	\subsubsection{The $\left(\omega-\omega^{\prime}\right)$- plane}
	The $(\omega-\omega')$-plane provides a crucial framework for studying the behavior of \textit{dark energy} and its impact on the evolution of the Universe. In this context, \( \omega \) represents the equation of state (EoS) parameter, while \( \omega' \) denotes the derivative of \( \omega \) with respect to the natural logarithm of the scale factor. This relationship is mathematically expressed as:
	
	\begin{equation}\label{e15}
		\omega'=\frac{\partial \omega}{\partial (lna)}
	\end{equation}
	The derivative \( \omega' \) provides essential information about the temporal evolution of \textit{dark energy}. Within the $(\omega-\omega')$-plane, the evolution of the Universe can typically be divided into four main phases: quintessence-like behavior, phantom-like behavior, cosmological constant-like behavior, and transient behavior.
	\begin{figure}[ht]
		\centering
		\includegraphics[width=9cm,height=6.5cm,angle=0]{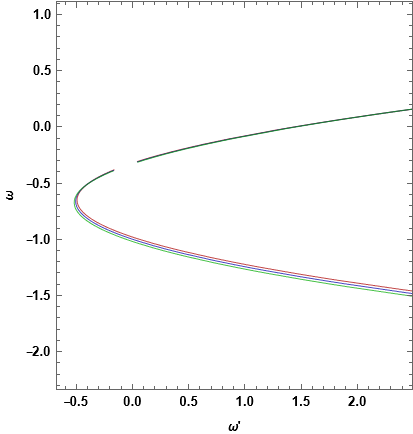}
		\caption{$\omega$ versus $\omega^{\prime}$}.\label{ww}
	\end{figure}  
	The \(w\)-\(w'\) plane in Fig. \ref{ww} illustrates the dynamical evolution of the EoS parameter \(\omega\) and its rate of change \(w'\) for different values of \(\eta\). 
	
	For \(\eta = 0.06\) (red curve), the universe starts in the matter-dominated phase with \(\omega > 0\), transitions into the quintessence phase (\(-1 < \omega < 0\)), and stabilizes near \(\omega = -1\). This indicates a freezing region in the interval \(-1 < \omega < -0.6\), where \(w'\) diminishes over time. 
	
	Similarly, for \(\eta = 0.07\) (blue curve), the universe follows a similar trajectory, approaching \(\Lambda\)CDM behavior as \(\omega\) stabilizes near \(-1\), with a freezing region in the interval \(-1 < \omega < -0.7\). 
	
	In contrast, for \(\eta = 0.08\) (green curve), the universe crosses the phantom divide line (\(\omega = -1\)) and enters the phantom phase (\(\omega < -1\)). This behavior corresponds to a thawing region in the interval \(-1.2 < \omega < -1\), where \(\omega\) evolves dynamically with an increasing \(w'\). Thus, smaller values of \(\eta\) result in freezing behavior, stabilizing \(\omega\) near \(-1\), while larger values of \(\eta\) lead to thawing behavior with \(\omega\) crossing into the phantom regime.

	\subsubsection{The energy conditions}
	In general relativity, energy requirements are essential limitations that guarantee the energy-momentum tensor, which depicts the distribution of matter and energy, complies with particular physical laws. These conditions are crucial in determining the evolution of the universe, especially in the context of \textit{dark energy} and modified gravity theories.
	
 According to the null energy condition (NEC), the energy-momentum tensor \( T_{\mu\nu} \) meets the inequality \( T_{\mu\nu}k^\mu k^\nu \geq 0 \) for any null vector \( k^\mu \). The dominant energy condition (DEC), which is more restrictive than the NEC, requires that for any future-directed timelike vector \( u^\mu \), the energy-momentum tensor satisfies \( T_{\mu\nu}u^\mu u^\nu \geq 0 \) and that \( T_{\mu\nu}u^\mu \) is future-directed. The strong energy condition (SEC) is the most stringent of the three, stating that for any future-directed timelike vector \( u^\mu \), the inequality \( (T_{\mu\nu} - \frac{1}{2}Tg_{\mu\nu})u^\mu u^\nu \geq 0 \) holds.
	
	These energy conditions influence various key aspects of the universe's dynamics, including cosmological expansion, the nature of \textit{dark energy}, black hole formation, and the development of cosmological singularities. They can be expressed mathematically in terms of energy density \( \rho \) and isotropic pressure \( p \) as follows:
	\begin{equation}\label{e17}
		\begin{split}
			& NEC: \rho + p \geq 0\\
			& DEC: \rho \geq 0 \;\; \text{and}\;\; \rho + p \geq 0 \\  
			& SEC: \rho + 3p \geq 0  \\  
		\end{split}  
	\end{equation}
	
	Thus, by considering the relevant energy conditions, we can evaluate the feasibility of our models. This approach will also aid in gaining a more realistic understanding of our universe.\\ 
	Null Energy Condition:\\
	The NEC is a crucial requirement in general relativity, asserting that $\rho + p \geq 0$. It represents a fundamental constraint on the energy-momentum tensor, ensuring that energy density measured by any observer is non-negative. 
	\begin{figure}[ht]
		\centering
		\includegraphics[scale=0.7]{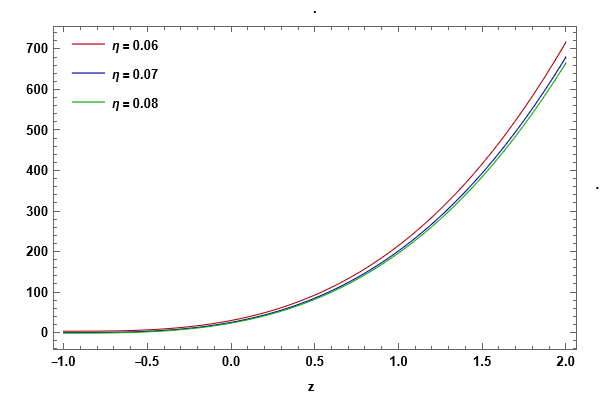}
		\caption{$(\rho+p)$ versus $z$}.\label{nec}
	\end{figure}  
	The corresponding figure \ref{nec}, illustrates the evolution of the NEC for three different values of $\eta : 0.06, 0.07, 0.08$. For each case, the curves generally satisfy the NEC over most redshifts, indicating that the underlying models are physically consistent and adhere to key principles of relativistic cosmology.\\
	Dominant Energy Condition:\\
	The DEC is a fundamental requirement in general relativity, which asserts that \(\rho - p \geq 0\). This condition ensures that the energy density is non-negative, preserving causality and the physical behavior of matter. 
	\begin{figure}[ht]
		\centering
		\includegraphics[scale=0.7]{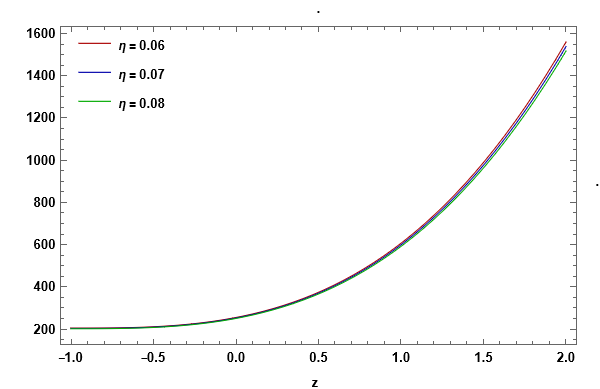}
		\caption{$(\rho-p)$ versus $z$}.\label{dec}
	\end{figure} 
	Physically, the satisfaction of the DEC implies that matter in the universe evolves in a manner consistent with standard relativistic frameworks, where violations of the DEC could lead to scenarios involving negative energy densities or repulsive gravitational effects. 
	
	The figure \ref{dec} displays three distinct curves corresponding to different values of \(\eta\): \(0.06\), \(0.07\), and \(0.08\). Across a wide range of redshifts, these curves often meet the DEC, demonstrating the models' physical consistency and adherence to basic energy constraints. Thus, the fulfillment of the DEC for these models further supports their feasibility in describing the realistic evolution of the universe.\\
	Strong Energy Condition:\\
	The SEC is a fundamental requirement in general relativity, stating that \(\rho + 3p \geq 0\). This condition reflects the idea that gravity should generally be attractive, as it ensures that the trace of the energy-momentum tensor contributes positively to the Ricci curvature, leading to deceleration in the cosmic expansion. Physically, the violation of the SEC is significant in modern cosmology, as it allows for the accelerated expansion of the universe. In standard cosmological models, the transition from a decelerating to an accelerating phase necessitates the breakdown of the SEC. The violation implies that the pressure of the dominant energy component (e.g., \textit{dark energy}) becomes sufficiently negative. This behavior aligns with current observational evidence, such as data from Type Ia supernovae and the cosmic microwave background, which confirm the universe's late-time acceleration. 
	\begin{figure}[ht]
		\centering
		\includegraphics[scale=0.7]{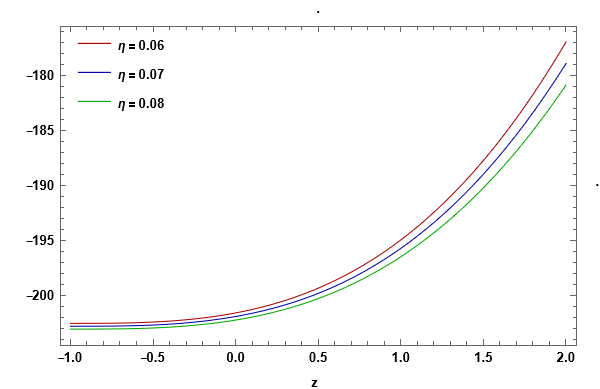}
		\caption{$(\rho+3p)$ versus $z$}.\label{sec}
	\end{figure} 
	The figure \ref{sec}, presents three distinct curves corresponding to different values of \(\eta\): \(0.06\), \(0.07\), and \(0.08\). It is observed that these curves generally violate the SEC at most redshifts, which suggests that the models are physically viable within the framework of an accelerating universe. Thus, the consistent violation of the SEC in these models supports their relevance in describing the observed dynamics of the universe, particularly its accelerated expansion.
	
	\section{Conclusion}
	The exploration of  gravity theory presented in this paper provides significant advancements in understanding the accelerated expansion of the universe and the enigmatic nature of dark energy. Unlike General Relativity (GR), which solely relies on the Einstein-Hilbert action, the  framework extends the geometric foundation of gravity by incorporating the Ricci scalar, the scalar parameter, and the trace of the energy-momentum tensor . This enriched theoretical structure enables a unified approach to cosmological modeling, addressing key phenomena such as the universe’s transition from deceleration to acceleration and the late-time dominance of dark energy.

Through the derivation and analysis of modified field equations, this study demonstrates the framework’s capacity to encompass various stages of cosmic evolution. The inclusion of observational constraints, such as those derived from OHD and SN Ia datasets, provides robust validation of the theoretical model. By obtaining consistent estimates for the Hubble constant  and matter energy density , the study reinforces the compatibility of  gravity with observational data. These results bridge the gap between theory and observation, affirming the reliability of the framework in capturing the universe’s dynamic nature.
The study’s findings on energy density, isotropic pressure, and the equation of state (EoS) parameter  further elucidate the behavior of cosmic fluids under the modified gravity framework. The negative evolution of isotropic pressure and its correlation with accelerated cosmic expansion align with the theoretical predictions of quintessence and phantom dark energy models. These results underscore the theory’s potential in addressing unresolved questions about the universe’s ultimate fate, including scenarios like the Big Rip.

The stability analysis, particularly through the squared sound velocity, confirms the theoretical model’s robustness under various cosmic conditions. Moreover, the evaluation of energy conditions, including the Null Energy Condition (NEC), Dominant Energy Condition (DEC), and Strong Energy Condition (SEC), provides additional insights into the physical consistency of the model. Notably, the violation of the SEC, consistent with late-time cosmic acceleration, aligns with the observed dynamics of the universe, bolstering the model’s relevance in contemporary cosmology.

In light of the findings of \cite{14a}, we acknowledge these developments and have expanded our analysis to consider models beyond the $\Lambda$CDM framework. By incorporating the parameter $f(R,\Sigma,T)$ gravity model, we allow for a dynamic equation of state for dark energy, accommodating potential deviations from a constant EoS. This approach aligns with recent observational trends and enhances the flexibility of our model in describing the universe's accelerated expansion and with \cite{14b}, our study adopts a more flexible approach by not assuming $\Lambda$CDM as the definitive model but rather as a limiting case within our broader framework. This perspective allows our model to accommodate potential deviations in the dark energy equation of state, aligning with recent DESI observations that question the constancy of this parameter. 

The  gravity framework’s ability to extend beyond the limitations of  and GR opens new avenues for theoretical exploration and observational testing. By introducing the scalar parameter , the model accommodates a broader spectrum of gravitational phenomena, enabling a more comprehensive analysis of cosmic evolution. This adaptability is crucial for addressing anomalies in current cosmological observations and refining predictions for future data.

	\section*{Declaration of competing interest}
	The authors state that none of the work described in this study could have been influenced by any known competing financial interests or personal relationships.
	
	\section*{Data availability}
The research presented in the paper did not use any data.
	
	\section*{Acknowledgments}
	The IUCAA, Pune, India, is acknowledged by the authors S. H. Shekh, \& A. Pradhan for giving the facility through the Visiting Associateship programmes. Additionally, the Science Committee of the Republic of Kazakhstan's Ministry of Science and Higher Education provided funding for this study (Grant No. AP23483654). Also, we sincerely appreciate the reviewers' valuable and thoughtful comments/feedback, which has significantly enhanced the clarity and strength of our study.

\end{document}